\documentclass[aps,prd,nofootinbib,showkeys,preprint,floatfix,showpacs]{revtex4}
\usepackage{amssymb}
\usepackage{amsmath}
\usepackage{amsfonts}
\usepackage{graphicx}
\usepackage{color}
\usepackage{wasysym}

\definecolor{lghtblue}{cmyk}{1,0.3,0,0}
\definecolor{midblue}{cmyk}{1,0.6,0,0}
\definecolor{mygold}{rgb}{0.67, 0.49, 0.04}       

\def\dsl{\Delta\!\!\!\!\backslash}

\newcommand{\AddrAHEP}{
  {\it AHEP Group, Instituto de F\'{\i}sica Corpuscular --
    C.S.I.C./Universitat de Val{\`e}ncia \\
    Edificio de Institutos de Paterna, Apartado 22085,
  E--46071 Val{\`e}ncia, Spain}}

\def\gsim{\raise0.3ex\hbox{$\;>$\kern-0.75em\raise-1.1ex\hbox{$\sim\;$}}}
\def\lsim{\raise0.3ex\hbox{$\;<$\kern-0.75em\raise-1.1ex\hbox{$\sim\;$}}}
\newcommand{\nn}{\nonumber}

\newcommand{\Da}{\Delta m^2_{\textsc{A}}}

\begin{document}

\preprint{IFIC/11-70}  

\title{Supersymmetric mass spectra and the seesaw type-I scale}

\author{C.~Arbel\'aez} \email{carolina@ific.uv.es}\affiliation{\AddrAHEP}

\author{M.~Hirsch} \email{mahirsch@ific.uv.es}\affiliation{\AddrAHEP}

\author{L.~Reichert} \email{reichert@ific.uv.es}\affiliation{\AddrAHEP}



\begin{abstract}
We calculate supersymmetric mass spectra with cMSSM boundary conditions 
and a type-I seesaw mechanism added to explain current neutrino data. 
Using published, estimated errors on SUSY mass observables for a combined 
LHC+ILC analysis, we perform a theoretical $\chi^2$ analysis to identify 
parameter regions where pure cMSSM and cMSSM plus seesaw type-I might 
be distinguishable with LHC+ILC data. The most important observables 
are determined to be the (left) smuon and selectron masses and the 
splitting between them, respectively. Splitting in the (left) 
smuon and selectrons is tiny in most of cMSSM parameter space, but can be 
quite sizeable for large values of the seesaw scale, $m_{SS}$. Thus, 
for very roughly $m_{SS} \ge 10^{14}$ GeV hints for type-I seesaw might 
appear in SUSY mass measurements. Since our numerical results depend 
sensitively on forecasted error bars, we discuss in some detail the 
accuracies, which need to be achieved, before a realistic analysis 
searching for signs of type-I seesaw in SUSY spectra can be carried 
out.

\end{abstract}

\maketitle


\section{Introduction}

The seesaw mechanism \cite{Minkowski:1977sc,seesaw,MohSen,Schechter:1980gr,%
Cheng:1980qt} provides a rationale for the observed smallness of neutrino 
masses \cite{Fukuda:1998mi,Ahmad:2002jz,Eguchi:2002dm,arXiv:0801.4589,
Schwetz:2008er,Nakamura:2010zzi}. However, due to the large mass scales 
involved, no direct experimental test of ``the seesaw'' will ever be 
possible. Extending the standard model (SM) only by a seesaw mechanism 
does not even allow for indirect tests, since all possible new observables 
are suppressed by (some power of) the small neutrino masses. 
\footnote{``Low-energy'' versions of the seesaw, such as inverse 
seesaw \cite{Mohapatra:1986bd} or linear seesaw \cite{Akhmedov:1995vm}, 
might allow for larger indirect effects. In this paper we will focus 
exclusively on the ``classical'' seesaw with a high (B-L) breaking scale.}

The situation looks less bleak in the supersymmetric version of the 
seesaw. This is essentially so, because soft SUSY breaking parameters 
are susceptible to all particles and couplings which appear in the 
renormalization group equation (RGE) running. Thus, assuming some 
simplified boundary conditions at an high energy scale, the SUSY 
softs at the electro-weak scale contain indirect information about 
all particles and intermediate scales. Perhaps the best known application 
of this idea is the example of lepton flavour violation (LFV) in seesaw 
type-I with cMSSM \footnote{``constrained'' Minimal Supersymmetric 
extension of the Standard Model, also sometimes called 
mSugra in the literature.} boundary conditions, discussed already 
in \cite{Borzumati:1986qx}. A plethora of papers on LFV, both for 
low-energy and for accelerator experiments, have been published 
since then (for an incomplete list see, for example, 
\cite{Hisano:1995nq,Hisano:1995cp,Ellis:2002fe,Deppisch:2002vz, 
Arganda:2005ji,Antusch:2006vw,Arganda:2007jw,Hisano:1998wn,
Blair:2002pg,Freitas:2005et,Petcov:2003zb,arXiv:0804.4072,929607}), 
most of them concentrating on seesaw type-I.

Seesaw type-I is defined as the exchange of fermionic singlets. 
At tree-level there is also the possibility to exchange (Y=2) 
scalar triplets \cite{Schechter:1980gr,Cheng:1980qt}, seesaw type-II, 
or exchange (Y=0) fermionic triplets, the so-called seesaw type-III 
\cite{Foot:1988aq,Ma:1998dn}. Common to all three seesaws is that for 
$m_{\nu} \sim \sqrt{\Da} \sim 0.05$ eV, where $\Da$ is the atmospheric 
neutrino mass splitting, and couplings of order ${\cal O}(1)$ the scale 
of the seesaw is estimated to be very roughly $m_{SS} \sim 10^{15}$ GeV. 
Much less work on SUSY seesaw type-II and type-III has been done than 
for type-I. For studies of LFV in SUSY seesaw type-II, see for example 
\cite{Rossi:2002zb,arXiv:0806.3361}, for type-III \cite{arXiv:1010.6000,
Biggio:2010me}. 

Apart from the appearance of LFV, adding a seesaw to the SM particle 
content also leads to changes in the absolute values of SUSY masses 
with respect to cMSSM expectations, at least in principle. Type-II 
and type-III seesaw add superfields, which are charged under the SM 
group. Thus, the running of the gauge couplings is affected, leading 
to potentially large changes in SUSY spectra at the EW scale. In 
\cite{Buckley:2006nv} it was pointed out, that for type-II and 
type-III seesaw certain combinations of soft SUSY breaking parameters 
are at 1-loop order nearly constant over large parts of cMSSM parameters 
space, but show a logarithmic dependence on $m_{SS}$. \footnote{These 
so-called invariants can be useful also in more  complicated models in 
which an inverse seesaw is embedded into an extended gauge group 
\cite{arXiv:1107.3412}.}  This was studied in more detail, including 
2-loop effects in the RGEs, for type-II in \cite{arXiv:0806.3361} and 
for type-III in \cite{arXiv:1010.6000}. Using forecasted errors on 
SUSY masses, obtained from full simulations \cite{Weiglein:2004hn,
AguilarSaavedra:2005pw}, the work \cite{Hirsch:2011cw} calculated the 
error with which the seesaw (type-II and -III) scale might be determined 
from LHC and future ILC \cite{AguilarSaavedra:2001rg} measurements. 
Interestingly, \cite{Hirsch:2011cw} concluded that, assuming cMSSM 
boundary conditions, ILC accuracies on SUSY masses should be sufficient 
to find at least some hints for a type-II/type-III seesaw, for practically 
all relevant values of the seesaw scale.

Seesaw type-I, on the other hand, adds only singlets. Changes in SUSY 
spectra are expected to be much smaller and, therefore, much harder to 
detect. Certainly because of this simple reasoning much fewer papers 
have studied this facet of the type-I SUSY seesaw so far. Running 
slepton masses with a type-I seesaw have been discussed qualitatively 
in \cite{Blair:2002pg,Freitas:2005et,Deppisch:2007xu,Kadota:2009sf}. 
In \cite{Abada:2010kj} it was discussed that in cMSSM extended 
by a type-I seesaw, splitting in the slepton sector can be considerably 
larger than in the pure cMSSM. This is interesting, since very small 
mass splittings in the smuon/selectron sector might be measurable 
at the LHC, if sleptons are on-shell in the decay chain $\chi^0_2 \to l^{\pm} 
{\tilde l}^{\mp} \to l^{\pm} l^{\mp}  \chi^0_1$ \cite{Allanach:2008ib}. 

In this paper, we calculate SUSY spectra with cMSSM boundary conditions 
and a seesaw type-I. We add three generations of right-handed neutrinos 
and take special care that observed neutrino masses and mixing angles are 
always correctly fitted. We then follow the procedure of \cite{Hirsch:2011cw}. 
Using predicted error bars on SUSY mass measurements for a combined LHC+ILC 
analysis, we construct fake ``experimental'' observables and use a 
$\chi^2$-analysis to estimate errors on the parameters of our model, most 
notably the seesaw scale. We identify regions in parameter space, where 
hints for a type-I seesaw might show up at the ILC/LHC and discuss 
{\em quantitatively} the accuracy which need to be achieved, before a 
realistic analysis searching for signs of type-I seesaw in SUSY spectra 
can be carried out.

The rest of this paper is organized as follows. In the next section 
we define the supersymetric seesaw type-I model, fix the notation 
and define the cMSSM. In section \ref{sect:numerics} we present our 
results. After a short discussion of the procedures and observables in 
section \ref{sect:prelim}, we show a simplified analysis, which allows 
to identify the most important observables and discuss their relevant 
errors in section \ref{sect:simpl}. Section \ref{sect:fullana} then 
shows our full numerical results. We then close with a short summary 
and discussion in section \ref{sect:cncl}.

\section{Setup}

\subsection{Supersymmetric seesaw type-I}
\label{sec:deftypeI}

In the case of seesaw type-I one postulates very heavy right-handed 
neutrinos with the following superpotential below the GUT scale, $M_{G}$:
\begin{eqnarray}
W_{I} &=& W_{MSSM} + W_{\nu}. \label{eq:superpotI}
\end{eqnarray}
Here $W_{MSSM}$ is the usual MSSM part and 
\begin{eqnarray}\label{eq:potSS}
 W_{\nu}& = &  {\widehat N}^c_i Y^\nu_{ij} {\widehat L}_j \cdot {\widehat H}_u
          + \frac{1}{2}{\widehat N}^c_i M_{R,ii}{\widehat N}^c_i \thickspace .
\end{eqnarray}
We have written eq. (\ref{eq:superpotI}) in the basis where $M_R$ and 
the charged lepton Yukawas are 
diagonal. In the seesaw one can always choose this basis without loss 
of generality. For the neutrino mass matrix, upon integrating out the 
heavy Majorana fields, one obtains the well-known {\em seesaw} formula
\begin{equation}
m_\nu = - \frac{v^2_u}{2} Y^{\nu,T} M^{-1}_R Y^\nu,
\label{eq:mnuI}
\end{equation}
valid up to order ${\cal O}(m_D/M_R)$, $m_D= \frac{v_u}{\sqrt{2}} Y^\nu$. 
Being complex symmetric, the light Majorana neutrino mass matrix in 
eq.~(\ref{eq:mnuI}), is diagonalized by a unitary $3\times 3$ matrix
  $U$~\cite{Schechter:1980gr}
\begin{equation}\label{diagmeff}
{\hat m_{\nu}} = U^T \cdot m_{\nu} \cdot U\ .
\end{equation}
Inverting the seesaw equation, eq.~(\ref{eq:mnuI}), allows to express 
$Y^{\nu}$ as \cite{Casas:2001sr}
\begin{equation}\label{Ynu}
Y^{\nu} =\sqrt{2}\frac{i}{v_u}\sqrt{\hat M_R}\cdot R \cdot \sqrt{{\hat
    m_{\nu}}} \cdot U^{\dagger},
\end{equation}
where the $\hat m_{\nu}$ and $\hat M_R$ are  diagonal matrices containing 
the corresponding eigenvalues. 
$R$ is  in general a complex orthogonal matrix. Note that, in the
special case $R={\bf 1}$, $Y^{\nu}$ contains only ``diagonal'' products
$\sqrt{M_im_{i}}$. For $U$ we will use the standard form
\begin{eqnarray}\label{def:unu}
U=
\left(
\begin{array}{ccc}
 c_{12}c_{13} & s_{12}c_{13}  & s_{13}e^{-i\delta}  \\
-s_{12}c_{23}-c_{12}s_{23}s_{13}e^{i\delta}  & 
c_{12}c_{23}-s_{12}s_{23}s_{13}e^{i\delta}  & s_{23}c_{13}  \\
s_{12}s_{23}-c_{12}c_{23}s_{13}e^{i\delta}  & 
-c_{12}s_{23}-s_{12}c_{23}s_{13}e^{i\delta}  & c_{23}c_{13}  
\end{array}
\right) 
 \times
 \left(
 \begin{array}{ccc}
 e^{i\alpha_1/2} & 0 & 0 \\
 0 & e^{i\alpha_2/2}  & 0 \\
 0 & 0 & 1
 \end{array}
 \right)
\end{eqnarray}
with $c_{ij} = \cos \theta_{ij}$ and $s_{ij} = \sin \theta_{ij}$. The
angles $\theta_{12}$, $\theta_{13}$ and $\theta_{23}$ are the solar
neutrino angle, the reactor angle and the atmospheric neutrino mixing 
angle, respectively. $\delta$ is the Dirac phase and $\alpha_i$ are 
Majorana phases. Since $U$ can be determined experimentally only 
up to an irrelevant overall phase, one can find different parameterizations 
of the Majorana phases in the literature.

Eq. (\ref{eq:mnuI}) contains 9 a priori unknown parameters, 
eq. (\ref{Ynu}) contains 18. The additional 9 unknowns encode 
the information about the high scale parameters, the three  
eigenvalues of $M_R$ and the 3 moduli and 3 phases of $R$.  

\subsection{cMSSM, type-I seesaw and RGEs}
\label{sec:defSugra}

The cMSSM is defined at the GUT-scale by: a common gaugino 
mass $M_{1/2}$, a common scalar mass $m_0$ and the trilinear coupling
$A_0$, which gets multiplied by the corresponding Yukawa couplings to
obtain the trilinear couplings in the soft SUSY breaking Lagrangian. 
In addition, at the electro-weak scale, $\tan\beta=v_u/v_d$ is fixed. 
Here, as usual, $v_d$ and $v_u$ are the vacuum expectation values (vevs) 
of the neutral component of $H_d$ and $H_u$, respectively.  Finally, 
the sign of the $\mu$ parameter has to be chosen.

Two-loop RGEs for general supersymmetric models have been given 
in \cite{Martin:1993zk}. \footnote{The only case not covered 
in \cite{Martin:1993zk} is models with more than one $U(1)$ gauge 
group. This case has been discussed recently in \cite{Fonseca:2011vn}.} 
In our numerical calculations we use {\tt SPheno3.1.5} \cite{Porod:2003um,
Porod:2011nf}, which solves the RGEs at 2-loop, including right-handed 
neutrinos. It is, however, useful for a qualitative understanding, to
consider first the simple solutions to the RGE for the slepton mass 
parameters found in the leading
log approximation \cite{Hisano:1995cp,Hisano:1998wn}, given by
\begin{eqnarray}\label{running}
(\Delta M_{\tilde L}^2)_{ij} & = &
 -\frac{1}{8\pi^2}(3 m_0^2 + A_0^2) 
  (Y^{\nu,\dagger}LY^{\nu})_{ij} \\ \nonumber
(\Delta A_l)_{ij} & = & -\frac{3}{8\pi^2}A_0Y_{l_i}
   (Y^{\nu,\dagger}LY^{\nu})_{ij}\\ \nonumber
(\Delta M_{\tilde E}^2)_{ij} & = & 0,
\end{eqnarray}
where only the parts proportional to the neutrino Yukawa couplings
have been written. The factor $L$ is defined as
\begin{equation}\label{deffacL}
L_{kl} = \log\Big(\frac{M_G}{M_{k}}\Big)\delta_{kl}. 
\end{equation}
Eq. (\ref{running}) shows that, within the type-I seesaw mechanism, 
the right slepton parameters do not run in the leading-log approximation.  
Thus, LFV is restricted to the sector of left-sleptons in
practice, apart from left-right mixing effects which could show up in
the scalar tau sector. Also note that for the trilinear parameters
running is suppressed by charged lepton masses.

It is important that the slepton mass-squareds involve a different 
combination of neutrino Yukawas and right-handed neutrino masses than 
the left-handed neutrino masses of eq.~(\ref{eq:mnuI}).  In fact, since 
$(Y^{\nu,\dagger}LY^{\nu})$ is a hermitian matrix, it obviously contains 
only nine free parameters \cite{Ellis:2002fe}, the same number of unknowns 
as on the right-hand side of eq.~(\ref{Ynu}), given that in principle all 
3 light neutrino masses, 3 mixing angles and 3 CP phases are potentially 
measurable.

Apart from the slepton mass matrices, $Y^{\nu}$ also enters the RGEs 
for $m_{H_u}^2$ at 1-loop level. However, we have found that the masses 
of the Higgs bosons are not very sensitive to the values of $Y^{\nu}$, 
see also next section. We thus do not give approximate expressions for 
$m_{H_u}^2$. For all other soft SUSY parameters, $Y^{\nu}$ enters only 
at the 2-loop level. Thus, the largest effects of the SUSY type-I seesaw 
are expected to be found in  the left slepton sector.

\section{Numerical results}
\label{sect:numerics}

\subsection{Preliminaries}
\label{sect:prelim}

We use {\tt SPheno3.1.5} \cite{Porod:2003um,Porod:2011nf} to 
calculate all SUSY spectra and fit the neutrino data. Unless 
noted otherwise the fit to neutrino data is done for strict 
normal hierarchy (i.e. $m_{\nu_1}=0$), best-fit values for the 
atmospheric and solar mass squared splitting \cite{Schwetz:2008er} 
and tri-bimaximal mixing angles \cite{hep-ph/0202074}. To reduce 
the number of free parameters in our fits, we assume right-handed 
neutrinos to be degenerate and $R$ to be the identity. The seesaw 
scale, called $m_{SS}$ below, is equal to the degenerate right-handed 
neutrino masses. We will comment on expected changes of our results, 
when any of these assumptions is dropped in the next subsections. 
Especially, recently there have been some indications for a non-zero 
reactor angle, both from the long-baseline experiment T2K 
\cite{arXiv:1111.0183} as well as from the first data in Double CHOOZ 
\cite{DC2011}. We will therefore comment also on non-zero 
values of $\theta_{13} = \theta_R$. 

SPheno solves the RGEs at 2-loop level and calculates the SUSY masses 
at 1-loop order, except for the Higgs mass, where the most important 
2-loop corrections have been implemented too. Theoretical errors 
in the calculation of the SUSY spectrum are thus expected to be 
much smaller than experimental errors at the LHC. However, since 
for the ILC one expects much smaller error bars, theory errors will 
become important at some point. We comment on theory errors in the 
discussion section.

Observables and their theoretically forecasted errors are taken from
the tables (5.13) and (5.14) of \cite{Weiglein:2004hn} and from
\cite{AguilarSaavedra:2005pw}. For the LHC we take into account the
``edge variables'': $(m_{ll})^{edge}$,
$(m_{lq})^{\text{edge}}_{\text{low}}$,
$(m_{lq})^{\text{edge}}_{\text{high}}$, $(m_{llq})_{\rm edge}$ and
$(m_{llq})_{\rm thresh}$ from the decay chain ${\tilde q}_L\to
\chi^0_2 q$ and $\chi^0_2 \to l {\tilde l} \to ll \chi^0_1$
\cite{Bachacou:1999zb,Allanach:2000kt,Lester:2001zx}. In addition, we
consider $(m_{llb})_{thresh}$, $(m_{\tau^+\tau^-})$ (from decays
involving the lighter stau) and the mass differences $\Delta_{{\tilde
g}{\tilde b}_i}= m_{\tilde g} - m_{{\tilde b}_i}$, with
$i = 1,2$, $\Delta_{{\tilde q}_R\chi^0_1}= m_{{\tilde q}_R} -
m_{\chi^0_1}$ and $\Delta_{{\tilde l}_L\chi^0_1}=
m_{{\tilde l}_L} - m_{\chi^0_1}$. Since $m_{{\tilde u}_R}\simeq m_{{\tilde
d}_R}\simeq m_{{\tilde c}_R}\simeq m_{{\tilde s}_R}$ applies for a
large range of the parameter space LHC measurements will not be able
to distinguish between the first two generation squarks. 
The combined errors for an LHC+ILC analysis, tables (5.14) of 
\cite{Weiglein:2004hn}, are dominated by the ILC for all non-coloured 
sparticles, except the stau. For us it is essential that both, left and 
right sleptons are within reach of the ILC. Also the two lightest 
neutralinos and the lighter chargino measured at ILC are important. 
The errors in 
\cite{Weiglein:2004hn} were calculated for relatively light 
SUSY spectra, thus we extrapolate them to our study points, see 
below, assuming constant relative errors on mass measurements. 
We will comment in some detail on the importance of this assumption 
below. Finally, we use the splitting in the selectron/smuon sector 
\cite{Allanach:2008ib} as an observable:

\begin{equation}\label{eq:defsplit}
\Delta(m_{\tilde e \tilde\mu}) = \frac{m_{\tilde e}-m_{\tilde \mu}}
{m_{\tilde l}^{mean}}.
\end{equation}
Here, ${m_{\tilde l}^{mean}} = \frac{1}{2}(m_{\tilde e}+m_{\tilde \mu})$. 
The LHC can, in principle, measure this splitting from the edge 
variables for both, left and right sleptons, if the corresponding 
scalars are on-shell. In cMSSM type-I seesaw only the left sector 
has a significant splitting, we therefore suppress the index ``L'' 
for brevity. For this splitting \cite{Allanach:2008ib} quote a 
``one sigma observability'' of 
$\Delta(m_{\tilde e \tilde\mu}) \sim 2.8$ $\permil$ for SPS1a.
\footnote{SPS1a has only the edge in the right-slepton sector on-shell, 
see discussion fig. (\ref{fig:chisqerr}).}  
For comparison, the errors on the left selectron and smuon 
mass at the ILC for this point are quoted as $\Delta(m_{\tilde e}) \simeq 1$ 
$\permil$ and $\Delta(m_{\tilde \mu}) \simeq 2.5$ 
$\permil$, respectively \cite{Weiglein:2004hn}.

The negative searches for SUSY by CMS \cite{arXiv:1109.2352} and 
ATLAS \cite{arXiv:1109.6572} define an excluded range in cMSSM  
parameter space, ruling out the lightest SPS study points, such as 
SPS1a' \cite{AguilarSaavedra:2005pw} or SPS3 \cite{Allanach:2002nj}. 
For our numerical study we define a set of five points, all of 
which are chosen to lie outside the LHC excluded region, but have 
the lightest non-coloured SUSY particles within reach of a 1 TeV 
linear collider. The points are defined as follows:
\begin{eqnarray}\label{eq:defp}\nn 
{\rm P}_1 &\rightarrow & (m_0=120, M_{1/2}=600, A_0=0, \tan\beta=10) \\ \nn
{\rm P}_2 &\rightarrow & (m_0=120, M_{1/2}=600, A_0=300, \tan\beta=10) \\ 
{\rm P}_3 &\rightarrow & (m_0=120, M_{1/2}=600, A_0=-300, \tan\beta=10) \\ \nn
{\rm P}_4 &\rightarrow & (m_0=180, M_{1/2}=550, A_0=0, \tan\beta=10) \\ \nn
{\rm P}_5 &\rightarrow & (m_0=180, M_{1/2}=550, A_0=300, \tan\beta=10) 
\end{eqnarray}
All points have $sgn(\mu)>0$, masses are in units of GeV. 
Points $P_1$-$P_3$ lie very close to the 
stau-coannihilation line. We have checked by an explicit calculation with 
MicrOmegas \cite{hep-ph/0405253,hep-ph/0607059,arXiv:0803.2360,arXiv:1004.1092}
that the relic density of the neutralino agrees with the current 
best fit value of $\Omega_{CDM}h^2$ within the quoted error 
bars \cite{Nakamura:2010zzi} for $P_1$. $P_4$ and $P_5$ have been 
chosen such that deviations from the pure cMSSM case are larger 
than in $P_1$-$P_3$, see eq.(\ref{running}),  i.e. to maximize the impact 
of the seesaw type-I on the spectra, see below.

\subsection{Observables and seesaw scale}
\label{sect:simpl}

In this subsection we will first keep all parameters at some fixed 
values, varying only the seesaw scale. These calculations are certainly 
simple-minded, but also very fast compared to the full Monte Carlo 
parameter scans, discussed later. However, as will be shown in the 
in the next subsection, there is nearly no correlation between different 
input parameters. Thus, the simple calculation discussed here already 
gives a quite accurate description of the results of the more complicated 
minimization procedures of the ``full'' calculation. Especially, this 
calculation allows us to identify the most important observables and 
discuss their maximally acceptable errors for our analysis.

\begin{figure}
 \centering
\includegraphics[height=50mm,width=70mm]{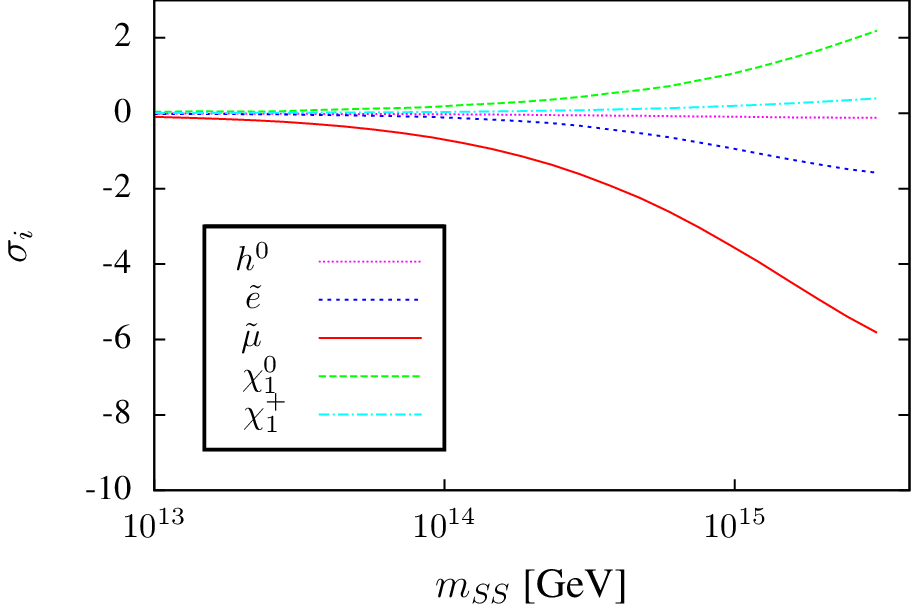} 
\includegraphics[height=50mm,width=70mm]{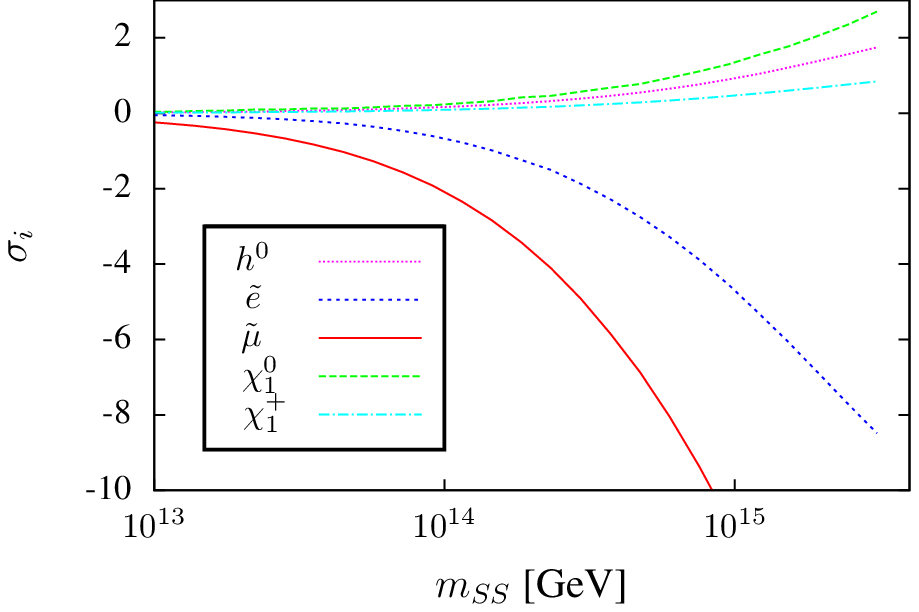}\\

\caption{\label{fig:sigall}Calculated deviations of masses from their 
nominal cMSSM values as function of $m_{SS}$ for the most important 
masses. To the left $P_1$, to the right $P_5$.}
\end{figure}

In fig. (\ref{fig:sigall}) we show 
\begin{equation}\label{eq:defsig}
\sigma_{i}= 
\frac{m_{i}^{m_{SS}}-m_{i}^{cMSSM}}{m_{i}^{cMSSM}}
\Big/\Delta(m_{i}),
\end{equation}
where $\Delta(m_{i})$ is the expected relative experimental error for 
the mass of sparticle i at the ILC, as a function of $m_{SS}$. We 
remind the reader that we assume that $\Delta(m_{i})$ can be extrapolated 
to our study points. To the 
left results for $P_1$ and to the right for $P_5$. $m_{i}^{cMSSM}$ is 
the value of the mass calculated in the cMSSM limit and $m_{i}^{m_{SS}}$ 
the corresponding mass for a seesaw scale of $m_{SS}$. These latter 
values have always been calculated fitting the Yukawa matrix of the 
neutrinos at $m_{SS}$, such that the best fit values of solar and 
atmospheric neutrino mass differences are obtained and $m_{\nu_1} \equiv 0$ 
is maintained. As expected the departures from the cMSSM values then 
increase with increasing seesaw scale. Note that the lines stop 
at values of $m_{SS} \sim (2-3) \times 10^{15}$ GeV, since for larger 
values neutrino Yukawas, which are required to fit the neutrino data, 
are non-perturbative. 

Significant departures with respect to the cMSSM values are found (with 
decreasing importance) for the following observables: 
left smuon mass, left selectron mass, mass of $\chi^0_1$, $m_{h^0}$ and 
$\chi^+_1$. We have checked that all other observables have much milder 
dependences on $m_{SS}$, as expected. The smuon mass is more 
important than the selectron mass, despite the latter having a smaller 
predicted error, due to our choice of degnerate right-handed neutrinos 
in the fits. With this assumption the running of the smuon mass has 
contributions from Yukawas responsible for both, atmospheric and solar 
scale, while the selectron has contributions from the Yukawas of the 
solar scale only. The change in $\chi^0_1$ and 
$\chi^+_1$ masses are small in absolute scale, but it is expected that 
ILC will measure these masses with very high accuracy. Also $m_{h^0}$ 
shows some mild dependence on $m_{SS}$, but on a scale of an expected 
experimental error of $50$ MeV 
\cite{AguilarSaavedra:2005pw}, i.e. much smaller than our current 
theoretical error, see below.

As the figure shows deviations from cMSSM expectations of the order 
of several standard deviations are reached for left smuon and 
selectron for values of $m_{SS}$ above $10^{14}$ GeV. Comparing the 
results for $P_1$ (left) with those for $P_5$ (right) it is confirmed 
that $P_5$ shows much larger deviations from cMSSM. We have checked that 
results for the other points $P_2$-$P_4$ fall in between the extremes 
of $P_1$ and $P_5$. Lines for $P_2$ and $P_3$ are nearly indistinguishable 
in such a plot, apart from some minor difference in the Higgs mass.

\begin{figure}
 \centering
\includegraphics[height=50mm,width=70mm]{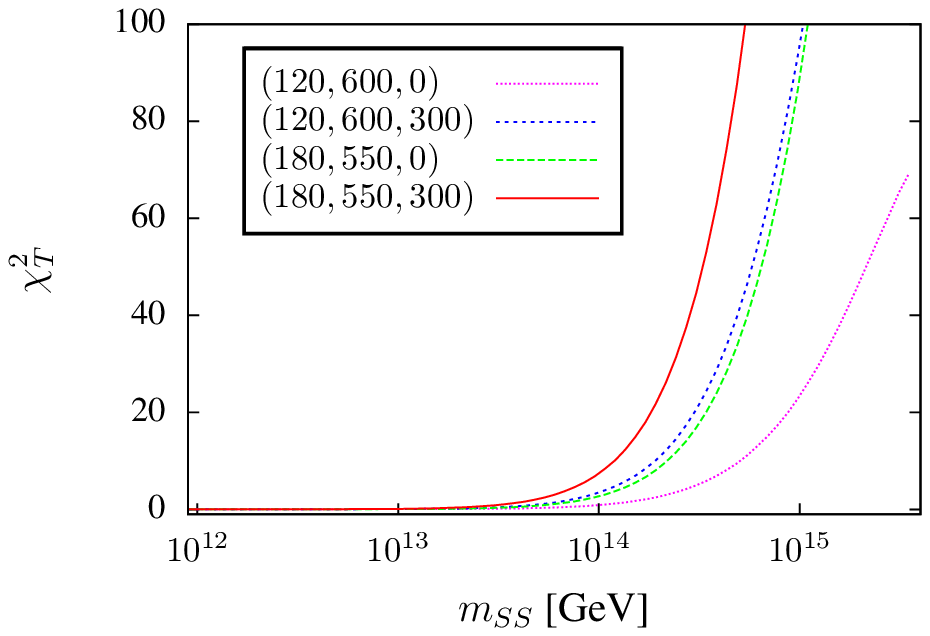}
\includegraphics[height=50mm,width=70mm]{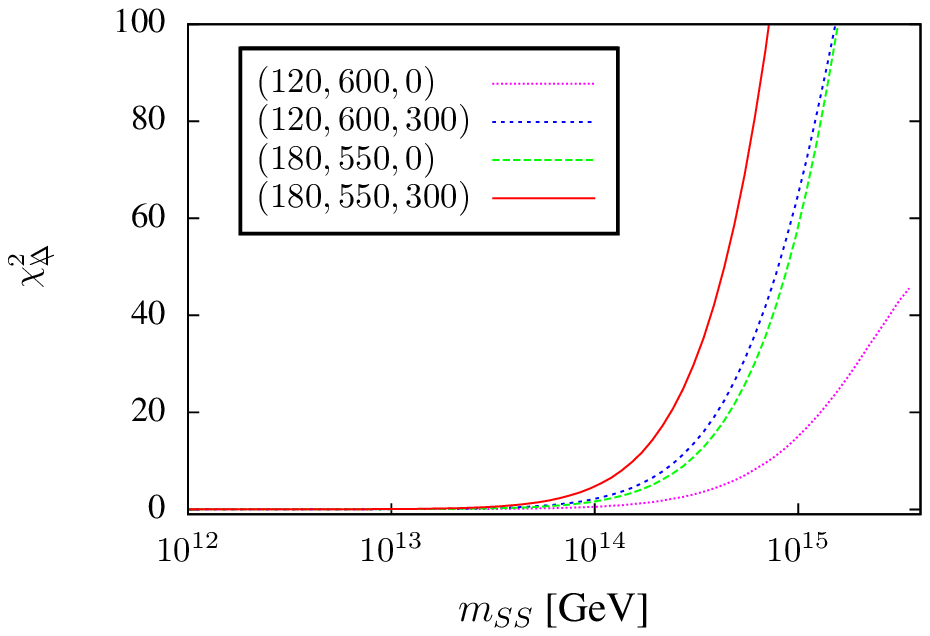}\\
\caption{\label{fig:chisq}Calculated $\chi^2$ as function of 
$m_{SS}$ for 4 different cMSSM points. To the left: Total $\chi^2$ 
including all observables, to the right total $\chi^2_{\dsl}$, i.e. 
$\chi_T$ without the mass splitting in the (left) smuon-selectron sector. 
Values quoted in the plots correspond to ($m_0,M_{1/2},A_0$). In 
all points shown we choose $\tan\beta=10$ and $\mu>0$. }
\end{figure}

In fig. (\ref{fig:chisq}) we show the calculated $\chi^2$ as a 
function of $m_{SS}$ for 4 different cMSSM points. Here, $\chi^2$ 
is calculated with respect to cMSSM expectations. To the left 
we show $\chi^2_T$ including all observables, to the right 
$\chi^2_T$ without the mass splitting in the (left) smuon-selectron 
sector. The figure demonstrates again that $P_1$ ($P_5$) has the 
smallest (largest) departures from cMSSM expectations. A non-zero 
value of $A_0$ can lead to significant departures from cMSSM 
expectations. Determination of $A_0$ from measurements 
involving 3rd generation sfermions and the lightest Higgs mass will 
therefore be important in fixing $m_{SS}$.

\begin{figure}
 \centering
\includegraphics[height=50mm,width=70mm]{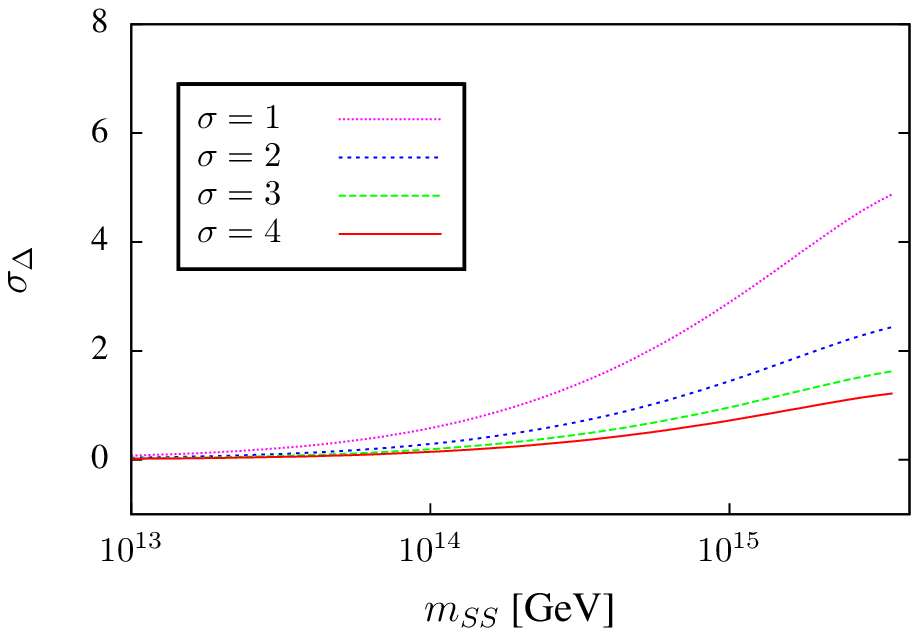}
\includegraphics[height=50mm,width=70mm]{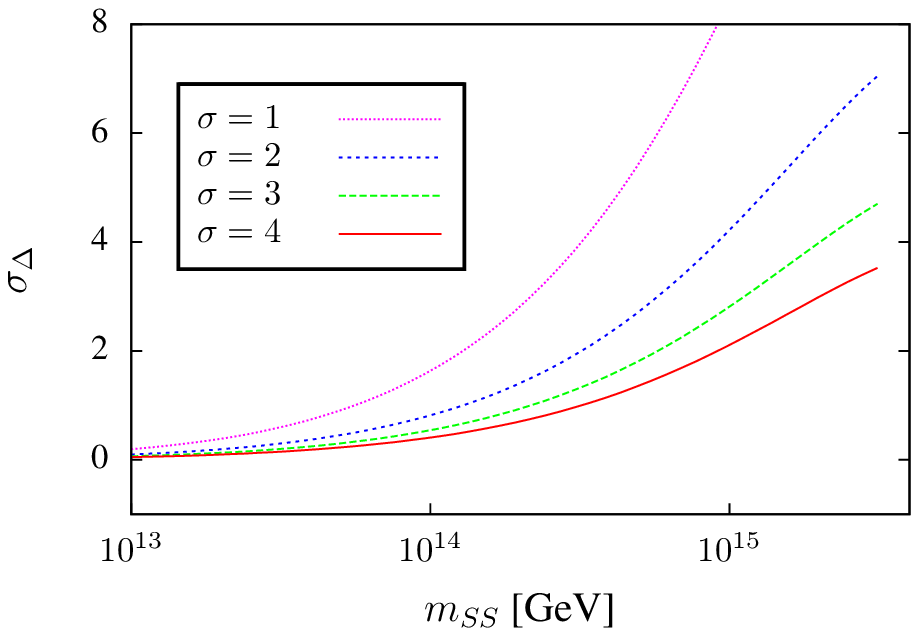}\\
\caption{\label{fig:chisqerr}Calculated $\chi^2$ for the observable 
$\Delta(m_{\tilde e \tilde\mu})$ as function of $m_{SS}$ for different 
values of its error. To the left: $P_1$; to the right $P_5$.}
\end{figure}

Fig. (\ref{fig:chisq}) also demonstrates that $\Delta(m_{\tilde e \tilde\mu})$ 
at its nominal error gives a significant contribution to the total $\chi^2$. 
Thus, LHC measurements only might already give some hints for a type-I 
seesaw \cite{Abada:2010kj}. However, with the rather large error bars 
of mass measurements at the LHC it will not be possible to fix the 
cMSSM parameters with sufficient accuracy to get a reliable error 
on the value of $m_{SS}$. Unfortunately, also the accuracy with which 
$\Delta(m_{\tilde e \tilde\mu})$ can be measured at the LHC is quite 
uncertain. According to \cite{Allanach:2008ib} such a splitting could 
be found for values as low as (few) $10^{-4}$ or as large as (several) 
percent, depending on the kinematical configuration realized in nature. 
Moreover, our points $P_1$-$P_5$ have heavier spectra than the ones 
studied in \cite{Allanach:2008ib}, so larger statistical errors are 
to be expected. 

Fig. (\ref{fig:chisqerr}) shows the relative deviation of 
$\Delta(m_{\tilde e \tilde\mu})$ for $P_1$ (left) and $P_5$ (right) 
for different assumed values of the error in this observable, relative
to cMSSM. Here, 
$\sigma=1,2,3,4$ means that we have multiplied the ``error'' quoted 
in  \cite{Allanach:2008ib} by factors $1,2,3,4$. The deviation drops 
below one sigma for any value of $m_{SS}$ shown for 
$P_1$ ($P_5$) when this error is larger than twice (six times) the 
nominal error. This implies that no hints for seesaw type-I can 
be found  in LHC data if the error on $\Delta(m_{\tilde e \tilde\mu})$ 
is larger than 5 $\permil$ (1.6 \%) in case of $P_1$ ($P_5$). 

We should also mention that the actual value of 
$\Delta(m_{\tilde e \tilde\mu})$ is not only a function of $m_{SS}$ 
and the cMSSM parameters, but also depends on the type of fit used 
to explain neutrino data. We have used degenerate right-handed neutrinos 
and $m_{\nu_1}\equiv 0$ in the plots shown above. Much smaller splittings 
are found for (a) nearly-degenerate light neutrinos, i.e. 
$m_{\nu_1}\ge 0.05$ eV; or (b) very hierarchical right-handed neutrinos. 
We have checked by an explicit calculation that, for example, for  $P_5$ 
and $m_{\nu_1}\equiv 0$, $\Delta \chi^2 \ge 5.89$ \footnote{$\Delta \chi^2 
\ge 5.89$ corresponds to 1 $\sigma$ c.l. for 5 free parameters.}
for values of $m_{SS}$ larger than $m_{SS} \simeq 1.6 \times 10^{14}$ GeV from 
$\Delta(m_{\tilde e \tilde\mu})$ alone, whereas the same $\Delta \chi^2$ 
is reached for $m_{\nu_1}= 0.05$ eV only for $m_{SS} \gsim 7 \times 10^{14}$ 
GeV. Consequently, even though one expects that a finite mass difference 
between left smuon and selectron is found in cMSSM type-I seesaw, this is 
by no means guaranteed.

Similar comments apply to the errors for the selectron and smuon mass 
at the ILC. For $P_1$ ($P_5$) the departure of the left selectron mass from 
the cMSSM expectations is smaller than 1 $\sigma$ even for $m_{SS} \sim 
3 \times 10^{15}$ if the error on this mass is larger than $1.5 \permil$ 
($1 \%$). For the left smuon the corresponding numbers are for $P_1$ 
 and $P_5$ approximately $1.5 \%$  and $5 \%$, respectively.

\begin{figure}
 \centering
\includegraphics[height=50mm,width=70mm]{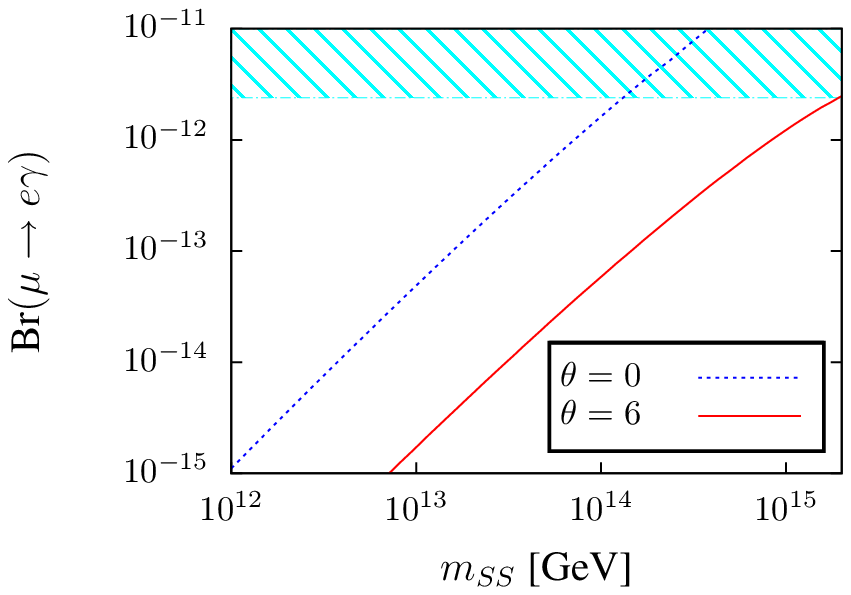}
\includegraphics[height=50mm,width=70mm]{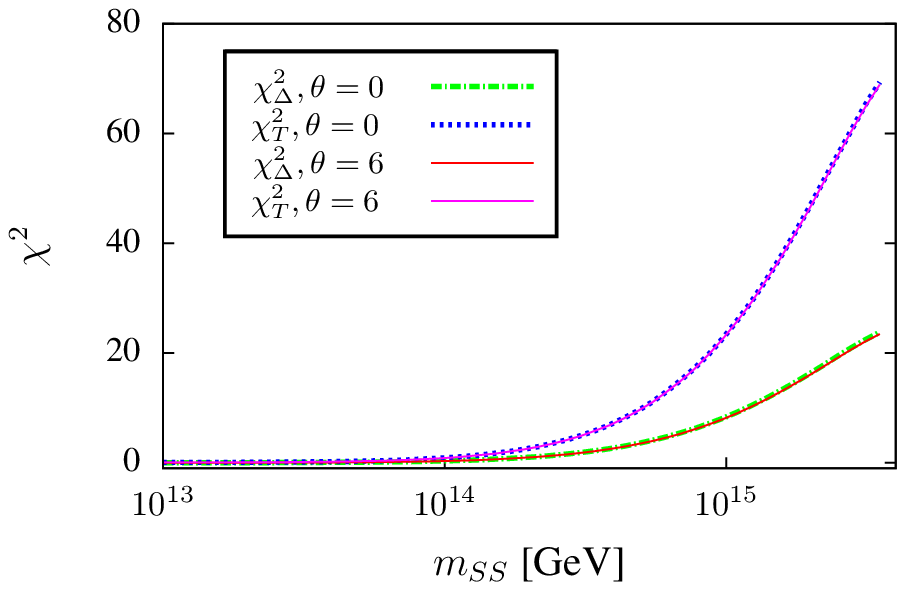}\\
\caption{\label{fig:chisqang}To the left Br($\mu\to e \gamma$) 
and to right calculated $\chi^2$ as function of $m_{SS}$ for two 
different values of the reactor angle $\theta_R$.}
\end{figure}

Naively one expects LFV violation to be large, whenever the neutrino 
Yukawa couplings are large, i.e. for large values of $m_{SS}$. That 
is, the regions testable by SUSY mass measurements could already 
be excluded by upper bounds on LFV, especially 
the recent upper bound on $\mu \to e \gamma$ by MEG \cite{arXiv:1107.5547}. 
That this conjecture is incorrect is demonstrated by the example shown in 
fig. (\ref{fig:chisqang}). In this figure we show the calculated 
Br($\mu \to e \gamma$) to the left and the calculated $\chi^2$ 
(total and only $\Delta(m_{\tilde e \tilde\mu})$) to the right 
for $\delta=\pi$ and two different values of the reactor angle, 
$\theta_{13}$ for the point $P_1$. For  $\theta_{13}=0$ all values 
of $m_{SS}$ above approximately $m_{SS} \sim 10^{14}$ GeV are excluded 
by the upper bound Br($\mu \to e \gamma$) $\le 2.4 \times 10^{-12}$ 
\cite{arXiv:1107.5547}. For $\theta_{13}=6^\circ$ nearly all 
values of $m_{SS}$ become allowed. At the same time, this ``small'' 
change in the Yukawas has practically no visible effect on the calculated 
$\chi^2$ from mass measurements as the plot on the right shows. This 
demonstrates that SUSY mass measurements and LFV probe different portions 
of seesaw type-I parameter space, contrary to what is sometimes claimed 
in the literature. That one can fit LFV and SUSY masses independently 
even for such a simple model as type-I seesaw is already obvious from 
eq. (\ref{running}): Even after fixing all low energy neutrino observables 
we still have nine unknown parameters to choose from to fit any entry of 
the left slepton masses {\em independently}.  

Fig. (\ref{fig:chisqang}) also shows that non-zero values of 
$\theta_{13}$, as preferred by the most recent experimental data 
\cite{arXiv:1111.0183,DC2011}, should have very little effect on our 
parameter scans. In our numerical scans, discussed next, we therefore 
keep $\theta_{13}=0$ unless mentioned otherwise. We will, however, also 
briefly comment on changes of our results, when $\theta_{13}$ is allowed 
to float within its current error.

\subsection{Numerical scans}
\label{sect:fullana}

For the determination of errors on the cMSSM parameters and $m_{SS}$ we 
have used two independent programmes, one based on MINUIT while the other 
uses a simple MonteCarlo procedure to scan over the free parameters. 
For a more detailed discussion see \cite{Hirsch:2011cw}. Plots shown 
below are obtained by the MonteCarlo procedure, but we have checked that 
results from MINUIT and our simplistic approach described above give 
very similar estimates for the $\chi^2$, with MINUIT only slightly 
improving the quality of the fit. In this section 
we always use all observables in the fits and quote all errors at 
1 $\sigma$ c.l., unless noted otherwise. Since our ``fake'' experimental 
data sets are perfect sets, the minimum of $\chi^2$ calculated equals 
zero and is thus not meaningful; only $\Delta\chi^2$ calculated with 
respect to the best fit points has any physical meaning in the 
plots shown below.

\begin{figure}
\includegraphics[height=50mm,width=70mm]{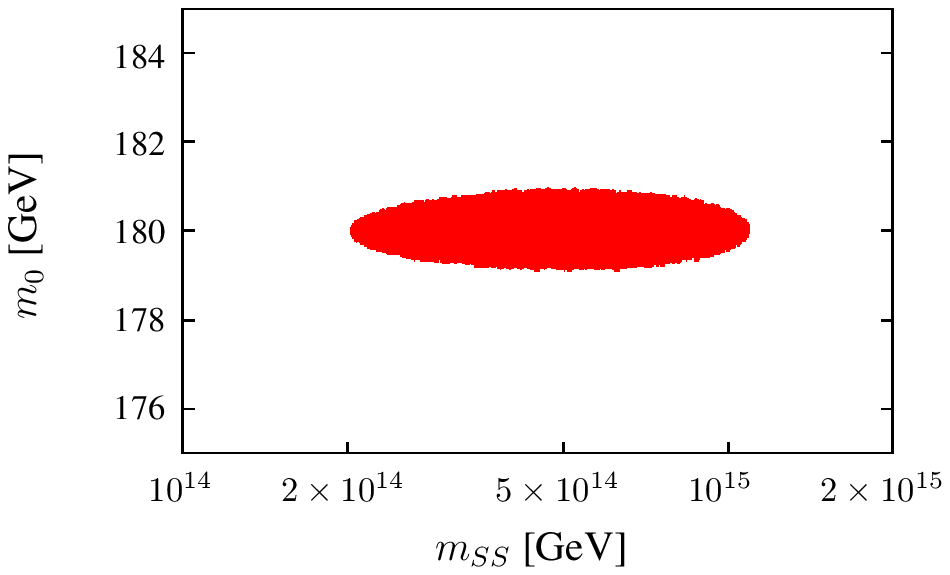}
\includegraphics[height=50mm,width=70mm]{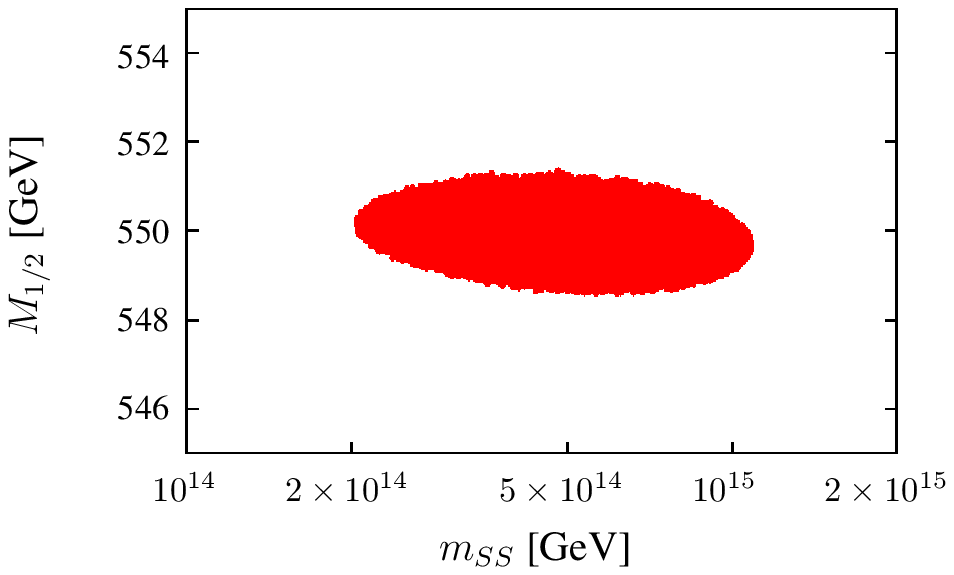}\\
\includegraphics[height=50mm,width=70mm]{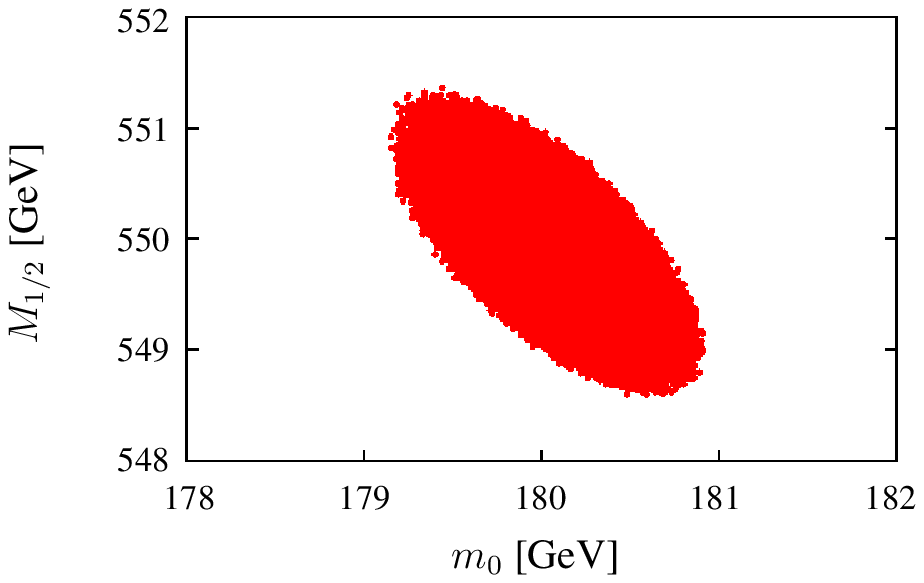}
\includegraphics[height=50mm,width=70mm]{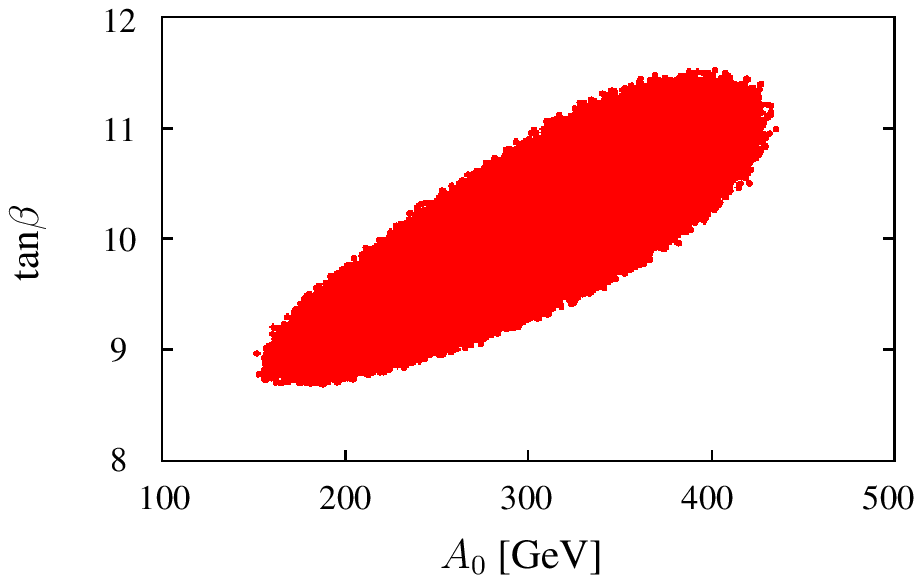}\\

\caption{\label{fig:paracorr}Calculated allowed parameter 
space for $m_0$, $M_{1/2}$, $\tan\beta$, $A_0$ and $m_{SS}$ for 
7 free parameters, $P_5$ and $m_{SS}=5 \times 10^{14}$ GeV. For discussion 
see text.}

\end{figure}

Fig. (\ref{fig:paracorr}) shows the allowed parameter space obtained in 
a MonteCarlo run for $m_0$, $M_{1/2}$, $\tan\beta$, $A_0$ and $m_{SS}$ for 
7 free parameters, $P_5$ and $m_{SS}=5 \times 10^{14}$ GeV. Shown are 
the allowed ranges of $m_0$ and $M_{1/2}$ versus $m_{SS}$, as well 
as $m_0$ versus $M_{1/2}$ and $\tan\beta$ versus $A_0$. On top 
of the 4 cMSSM parameters and $m_{SS}$ in this calculation we allow 
the solar angle ($\theta_{12}$) and the atmospheric angle ($\theta_{23}$) 
to float freely within their allowed range. Errors on 
neutrino angles for this plot are taken from \cite{arXiv:1103.0734}. 
Plots for other points 
and/or different sets of free parameters look qualitatively very 
similar to the example shown in the figure. There is very little 
correlation among different parameters, contrary to the situation found 
in case of seesaw type-II and type-III \cite{Hirsch:2011cw}. Especially 
no correlations between $m_0$, $M_{1/2}$ and $m_{SS}$ are found. 
However, there is some correlation between $\tan\beta$ and $A_0$, 
driven by the fact that $m_{h^0_1}$ alone can only fix a certain 
combination of these two parameters well. The correlation 
between $\tan\beta$ and $A_0$ is slightly stronger than in the 
cMSSM case, due to the contribution of $A_0$ in the running of 
slepton masses, see eq. (\ref{running}).

For our assumed set of measurements, $m_0$ and $M_{1/2}$ are mainly 
determined by the highly accurate measurements of right slepton and 
gaugino masses of the ILC. $A_0$ and $\tan\beta$ are fixed by a 
combination of the lightest Higgs mass and the lighter stau mass. 
LHC measurements help to break degeneracies in parameter space, but are 
much less important. We stress that the highly accurate determination of 
cMSSM parameters shown in fig. (\ref{fig:paracorr}) is a prerequisite 
for determining reliable errors on $m_{SS}$. \footnote{We have checked 
this explicitly in a calculation using only LHC observables.}

\begin{figure}
\includegraphics[height=50mm,width=70mm]{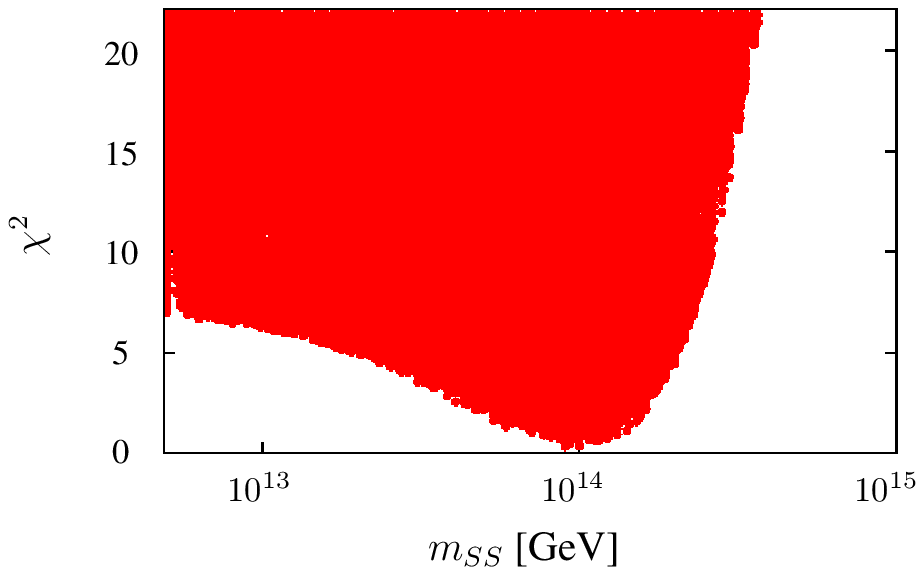}
\includegraphics[height=50mm,width=70mm]{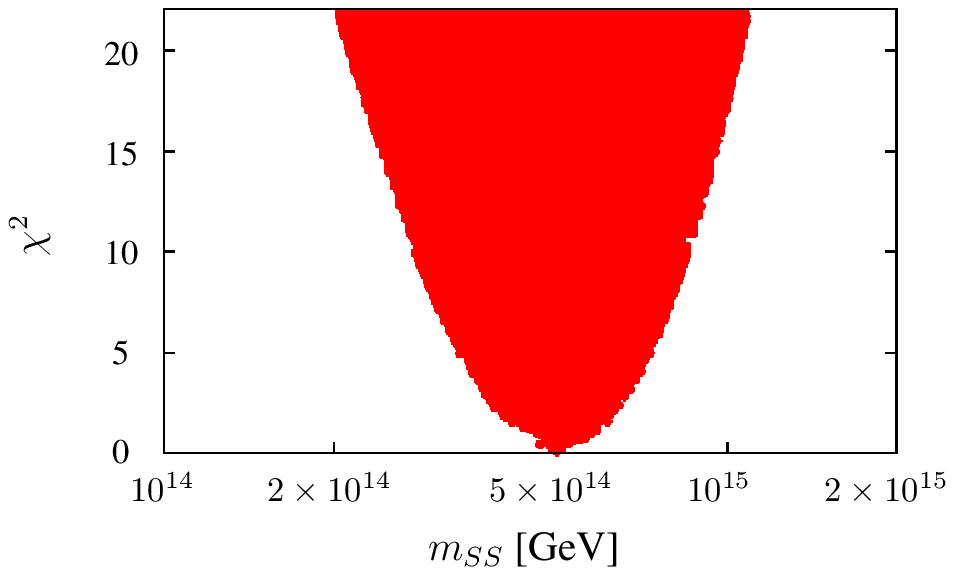}\\
\caption{\label{fig:chivmss}Calculated $\chi^2$ distribution versus 
$m_{SS}$ for 7 free parameters, $P_5$ and $m_{SS}= 10^{14}$ GeV 
(to the left) and  $m_{SS}=5 \times 10^{14}$ GeV (to the right).}
\end{figure}

Fig. (\ref{fig:chivmss}) shows calculated $\chi^2$ distributions versus 
$m_{SS}$ for the same 7 free parameters as in fig. (\ref{fig:paracorr}), 
$P_5$ and $m_{SS}= 10^{14}$ GeV (to the left) and  $m_{SS}=5 \times 10^{14}$ 
GeV (to the right). For the latter an upper (lower) limit of 
$m_{SS} \simeq 8 \times 10^{14}$ GeV ($m_{SS} \simeq 3\times  10^{14}$ GeV) 
is found. For $m_{SS}= 10^{14}$ GeV a clear upper limit is found, but 
for low values of $m_{SS}$ the $\chi^2$ distribution flattens out 
at $\Delta\chi^2 \sim 6.5$. This different behaviour can be understood 
with the help of the results of the previous subsection, see fig. 
(\ref{fig:chisq}). For $m_{SS}=5 \times 10^{14}$ GeV, there exists a 
notable difference in some observables with respect to the cMSSM 
expectation, especially left smuon and selectron mass can no longer 
be adequately fitted by varying $m_0$ and $M_{1/2}$ alone, without 
destroying the agreement with ``data'' for right sleptons and gauginos. 
Therefore both, a lower and an upper limit on $m_{SS}$ exist for this 
point. The situation is different for $m_{SS}= 10^{14}$ GeV, for which 
the spectrum is much closer to cMSSM expectations. Larger values of 
$m_{SS}$ are excluded, since they would require larger Yukawas, i.e. 
larger deviation 
from cMSSM than observed. Smaller values of $m_{SS}$, on the other hand, 
have ever smaller values of $Y^{\nu}$, i.e. come closer and closer 
to cMSSM expectations. For an input value of $m_{SS}$ just below 
$m_{SS}= 10^{14}$ GeV there is then no longer any lower limit on 
$m_{SS}$, i.e. the data becomes perfectly consistent with a pure 
cMSSM calculation. In this case one can only ``exclude'' a certain 
range of the seesaw, say values of $m_{SS}$ above a few $10^{14}$ 
GeV.

\begin{figure}
\includegraphics[height=50mm,width=70mm]{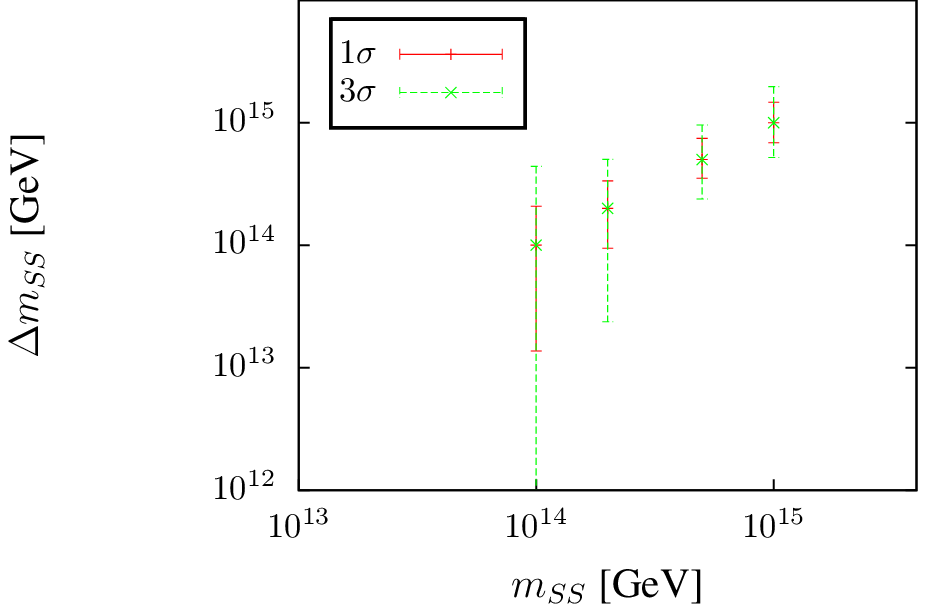}
\includegraphics[height=50mm,width=70mm]{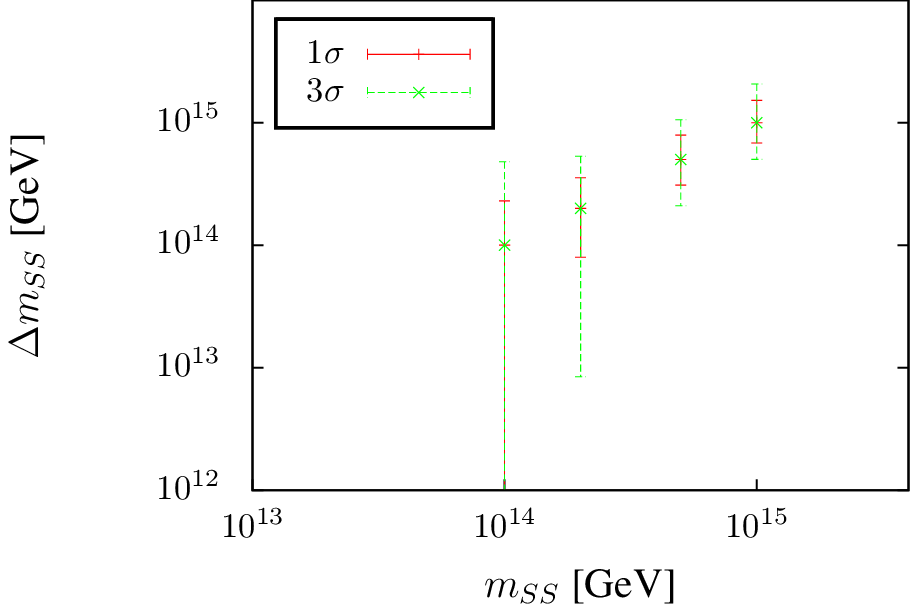}\\
\caption{\label{fig:1a3sig}Calculated allowed range of $m_{SS}$ 
versus $m_{SS}$ for 5 (left) and 7 (right) free parameters and $P_5$. 
The two different error bars correspond to 1 and 3 $\sigma$ c.l.}

\end{figure}

One standard deviation is, of course, too little to claim an observation. 
We therefore show in fig. (\ref{fig:1a3sig}) $\Delta(m_{SS})$ versus 
$m_{SS}$ for 5 (left) and 7 (right) free parameters and $P_5$ at 1 
and 3 $\sigma$ c.l. 
At $m_{SS}=10^{14}$ formally a 1 sigma ``evidence'' could be reached, 
but at 3 $\sigma$ c.l. the spectrum is perfectly consistent with a 
pure cMSSM. For larger values of $m_{SS}$, however, several standard 
deviations can be reached. For the two largest values of $m_{SS}$ 
calculated in this figure, a 5 $\sigma$ ``discovery'' is possible.

Fig. (\ref{fig:1a3sig}) shows $\Delta(m_{SS})$ for 5 and 7 free parameters.  
We have repeated this exercise for different sets of free parameters 
and $m_{SS}=5 \times 10^{14}$. Here, 5 free parameters correspond to 
the 4 cMSSM parameters plus $m_{SS}$, 7 free parameters 
are the original 5 plus $\theta_{12}$ and $\theta_{23}$. 
We have also tried other combinations such as 6 parameters: original 
5 plus $\theta_R$ and 8 parameters, where we let all 3 neutrino angles 
float freely. Sets with larger 
numbers of free parameters are no longer sufficiently sampled in our 
MonteCarlo runs, so we do not give numbers for these, although in 
principle the calculation could allow also to let the neutrinos mass 
squared differences to float freely. Error bars are slightly larger 
for larger number of free parameters, as expected. However, since there 
is little or no correlation among the parameters, the differences 
are so small as to be completely irrelevant.

\section{Summary and discussion}
\label{sect:cncl}

We have discussed the prospects for finding indirect hints for type-I 
seesaw in SUSY mass measurements. Since type-I seesaw adds only singlets 
to the SM particle content, only very few observables are affected and 
all changes in masses are small, even in the most favourable circumstances. 
Per-mille level accuracies will be needed, i.e. measurements at an ILC, 
before any quantitative attempt searching for type-I seesaw can hope for 
success, even assuming admittedly simplistic cMSSM boundary condtions. 

Our calculation confirms {\em quantitatively} that slepton mass measurements 
can contain information about the type-I seesaw. Right sleptons are 
expected to be degenerate, while the left smuon and selectron show a 
potentially measurable splitting between their masses. If such a situation 
is indeed found, an estimate of $m_{SS}$ might be derivable from ILC 
SUSY mass measurements. 

Above we have commented only on experimental errors. However, given the 
per-mille requirements on accuracy, stressed several times, also 
theoretical errors in the calculation of SUSY spectra are important. 
Various potential sources of errors come to mind. First of all, 
a 1-loop calculation of SUSY masses is almost certainly not accurate 
enough for our purposes. We have tried to estimate the importance of 
higher loop orders, varying the renormalization scale in the numerical 
calculation. Changes of smuon and selectron mass found are of the order 
of the ILC error or even larger, depending on SUSY point and variation of 
scale. For the mass of the lightest Higgs boson it has been shown 
that even different calculations at 2-loop still disagree at a level 
of few GeV \cite{hep-ph/0406166}. Second, our calculation assumes 
a perfect knowledge of the GUT scale. Changes in the GUT scale do 
lead to sizeable changes in the calculated spectra for the same 
cMSSM parameters, which can be easily of the order of the required 
precision of the calculation and larger. In this sense, 
$\Delta(m_{\tilde e \tilde\mu})$ is an especially nice observable, 
since here the GUT scale uncertainty nearly cancels out in the 
calculation. In summary, if ILC accuracies on SUSY masses can 
indeed be reached experimentally, progress on the theoretical side 
will become necessary too.

In our calculations, we have considered only SUSY masses. We have 
not taken into account data from lepton flavour violation, mainly 
because currently only upper limits are available. If in the future 
finite values for $\l_i \to \l_j + \gamma$ become available, it 
would be very interesting to see, how much could be learned about 
the type-I seesaw parameters in a combined fit. Including LFV 
one could maybe also allow for non-degenerate right-handed neutrinos 
in the fits. 

And, finally, despite all the limitations of our study, we find it 
very encouraging that hints for type-I seesaw might be found in SUSY 
mass measurements at all. We stress again, that LFV and SUSY mass 
measurements test different portions of seesaw parameter space. For 
a more complete ``reconstruction'' of seesaw parameters, than what we 
have attempted here, both kinds of measurements would be needed.

\section*{Acknowledgments}

We thank W. Porod for his patient help, discussing many detailed 
aspects of the numerics of SPheno during the course of this work. 
This work was supported by the Spanish MICINN under grants
FPA2008-00319/FPA and FPA2011-22975 by the MULTIDARK Consolider 
CSD2009-00064, by the Generalitat Valenciana grant Prometeo/2009/091 
and by the EU grant UNILHC PITN-GA-2009-237920. C.A. thanks the 
Generalitat Valenciana for support through the GRISOLIA programme.


\begin{thebibliography}{10}

\bibitem{Minkowski:1977sc}
  P.~Minkowski,
  Phys.\ Lett.\ B {\bf 67} (1977) 421.

\bibitem{seesaw}
T.~Yanagida, in {\it KEK lectures}, ed.  O.~Sawada and A.~Sugamoto,
KEK, 1979;
M Gell-Mann, P Ramond, R. Slansky, in {\it Supergravity}, ed. P. van
Niewenhuizen and D. Freedman (North Holland, 1979);

\bibitem{MohSen}
R.N.~Mohapatra and G.~Senjanovic, {\sl Phys. Rev. Lett.}\/ {\bf 44}
912 (1980).

\bibitem{Schechter:1980gr}
  J.~Schechter and J.~W.~F.~Valle,
  Phys.\ Rev.\ D {\bf 22}, 2227 (1980).

\bibitem{Cheng:1980qt}
  T.~P.~Cheng and L.~F.~Li,
  Phys.\ Rev.\  D {\bf 22}, 2860 (1980).


\bibitem{Fukuda:1998mi}
Y.~Fukuda {\it et al.}  [Super-Kamiokande Collaboration],
Phys.\ Rev.\ Lett.\  {\bf 81}, 1562 (1998)
\bibitem{Ahmad:2002jz}
SNO, Q.~R. Ahmad {\em et~al.},
\newblock Phys. Rev. Lett. {\bf 89}, 011301 (2002), [nucl-ex/0204008].
\bibitem{Eguchi:2002dm}
KamLAND, K.~Eguchi {\em et~al.},
\newblock Phys. Rev. Lett. {\bf 90}, 021802 (2003), [hep-ex/0212021].

\bibitem{arXiv:0801.4589} 
  S.~Abe {\it et al.} [KamLAND Collaboration],
  Phys.\ Rev.\ Lett.\ \ {\bf 100}, 221803  (2008)
  [arXiv:0801.4589 [hep-ex]].



\bibitem{Schwetz:2008er}
  For a recent review on the status of neutrino oscillation data, see: 
  T.~Schwetz, M.~A.~Tortola and J.~W.~F.~Valle,
  New J.\ Phys.\  {\bf 10}, 113011 (2008)
  [arXiv:0808.2016 [hep-ph]]. 
  Version 3 on the arXive is updated with data until Feb 2010 

\bibitem{Nakamura:2010zzi}
  K.~Nakamura {\it et al.}  [Particle Data Group],
  J.\ Phys.\ G {\bf 37} (2010) 075021.

\bibitem{Mohapatra:1986bd}
R.~N. Mohapatra and J.~W.~F. Valle,
\newblock Phys. Rev. {\bf D34}, 1642 (1986).

\bibitem{Akhmedov:1995vm}
E.~Akhmedov, M.~Lindner, E.~Schnapka and J.~W.~F. Valle,
\newblock Phys. Rev. {\bf D53}, 2752 (1996), [hep-ph/9509255];
\newblock Phys. Lett. {\bf B368}, 270 (1996), [hep-ph/9507275].


\bibitem{Borzumati:1986qx}
  F.~Borzumati and A.~Masiero,
  Phys.\ Rev.\ Lett.\  {\bf 57}, 961 (1986).

\bibitem{Hisano:1995nq}
  J.~Hisano, T.~Moroi, K.~Tobe, M.~Yamaguchi and T.~Yanagida,
  Phys.\ Lett.\  B {\bf 357}, 579 (1995)
  [arXiv:hep-ph/9501407].

\bibitem{Hisano:1995cp}
  J.~Hisano, T.~Moroi, K.~Tobe and M.~Yamaguchi,
  Phys.\ Rev.\  D {\bf 53}, 2442 (1996)
  [arXiv:hep-ph/9510309].

\bibitem{Ellis:2002fe}
  J.~R.~Ellis, J.~Hisano, M.~Raidal and Y.~Shimizu,
  Phys.\ Rev.\  D {\bf 66}, 115013 (2002)
  [arXiv:hep-ph/0206110].

\bibitem{Deppisch:2002vz}
  F.~Deppisch, H.~Pas, A.~Redelbach, R.~Ruckl and Y.~Shimizu,
  Eur.\ Phys.\ J.\  C {\bf 28}, 365 (2003)
  [arXiv:hep-ph/0206122].

\bibitem{Arganda:2005ji}
  E.~Arganda and M.~J.~Herrero,
  Phys.\ Rev.\  D {\bf 73}, 055003 (2006)
  [arXiv:hep-ph/0510405].

\bibitem{Antusch:2006vw}
  S.~Antusch, E.~Arganda, M.~J.~Herrero and A.~M.~Teixeira,
  JHEP {\bf 0611}, 090 (2006)
  [arXiv:hep-ph/0607263].

\bibitem{Arganda:2007jw}
  E.~Arganda, M.~J.~Herrero and A.~M.~Teixeira,
  JHEP {\bf 0710}, 104 (2007)
  [arXiv:0707.2955 [hep-ph]].

\bibitem{Hisano:1998wn}
  J.~Hisano, M.~M.~Nojiri, Y.~Shimizu and M.~Tanaka,
  Phys.\ Rev.\  D {\bf 60}, 055008 (1999)
  [arXiv:hep-ph/9808410].

\bibitem{Blair:2002pg}
G.~A. Blair, W.~Porod and P.~M. Zerwas,
\newblock Eur. Phys. J. {\bf C27}, 263 (2003), [hep-ph/0210058].

\bibitem{Freitas:2005et}
A.~Freitas, W.~Porod and P.~M. Zerwas,
\newblock Phys. Rev. {\bf D72}, 115002 (2005), [hep-ph/0509056].

\bibitem{Petcov:2003zb}
  S.~T.~Petcov, S.~Profumo, Y.~Takanishi and C.~E.~Yaguna,
  Nucl.\ Phys.\  B {\bf 676} (2004) 453
  [arXiv:hep-ph/0306195];
  S.~Pascoli, S.~T.~Petcov and C.~E.~Yaguna,
  Phys.\ Lett.\  B {\bf 564} (2003) 241
  [arXiv:hep-ph/0301095];
  S.~T.~Petcov, T.~Shindou and Y.~Takanishi,
  Nucl.\ Phys.\  B {\bf 738} (2006) 219
  [arXiv:hep-ph/0508243];
  S.~T.~Petcov and T.~Shindou,
  Phys.\ Rev.\  D {\bf 74} (2006) 073006
  [arXiv:hep-ph/0605151].

\bibitem{arXiv:0804.4072} 
  M.~Hirsch, J.~W.~F.~Valle, W.~Porod, J.~C.~Romao and A.~Villanova del Moral,
  Phys.\ Rev.\ D\ {\bf 78}, 013006  (2008)
  [arXiv:0804.4072 [hep-ph]].

\bibitem{929607} 
  M.~Hirsch,
  Nucl.\ Phys.\ Proc.\ Suppl.\ \ {\bf 217}, 318  (2011).

\bibitem{Foot:1988aq}
  R.~Foot, H.~Lew, X.~G.~He and G.~C.~Joshi,
  Z.\ Phys.\  C {\bf 44}, 441 (1989).

\bibitem{Ma:1998dn}
  E.~Ma,
  Phys.\ Rev.\ Lett.\  {\bf 81}, 1171 (1998)
  [arXiv:hep-ph/9805219].

\bibitem{Rossi:2002zb}
  A.~Rossi,
  Phys.\ Rev.\  D {\bf 66}, 075003 (2002)
  [arXiv:hep-ph/0207006].

\bibitem{arXiv:0806.3361} 
  M.~Hirsch, S.~Kaneko and W.~Porod,
  Phys.\ Rev.\ D\ {\bf 78}, 093004  (2008)
  [arXiv:0806.3361 [hep-ph]].


\bibitem{arXiv:1010.6000} 
  J.~N.~Esteves, J.~C.~Romao, M.~Hirsch, F.~Staub and W.~Porod,
  Phys.\ Rev.\ D\ {\bf 83}, 013003  (2011)
  [arXiv:1010.6000 [hep-ph]].

\bibitem{Biggio:2010me}
  C.~Biggio and L.~Calibbi,
  arXiv:1007.3750 [hep-ph].


\bibitem{Buckley:2006nv}
M.~R.~Buckley and H.~Murayama,
Phys.\ Rev.\ Lett.\  {\bf 97}, 231801 (2006)
[arXiv:hep-ph/0606088].

\bibitem{arXiv:1107.3412} 
  V.~De Romeri, M.~Hirsch and M.~Malinsky,
  Phys.\ Rev.\ D\ {\bf 84}, 053012  (2011)
  [arXiv:1107.3412 [hep-ph]].

\bibitem{Weiglein:2004hn}
  G.~Weiglein {\it et al.}  [LHC/LC Study Group],
  Phys.\ Rept.\  {\bf 426} (2006) 47
  [arXiv:hep-ph/0410364].

\bibitem{AguilarSaavedra:2005pw}
  J.~A.~Aguilar-Saavedra {\it et al.},
  Eur.\ Phys.\ J.\  C {\bf 46}, 43 (2006)
  [arXiv:hep-ph/0511344].


\bibitem{Hirsch:2011cw}
  M.~Hirsch, L.~Reichert, W.~Porod,
  JHEP {\bf 1105}, 086 (2011).
  [arXiv:1101.2140 [hep-ph]].

\bibitem{AguilarSaavedra:2001rg}
  J.~A.~Aguilar-Saavedra {\it et al.}  [ECFA/DESY LC Physics Working Group],
  arXiv:hep-ph/0106315.



\bibitem{Deppisch:2007xu}
  F.~Deppisch, A.~Freitas, W.~Porod and P.~M.~Zerwas,
  Phys.\ Rev.\  D {\bf 77} (2008) 075009
  [arXiv:0712.0361 [hep-ph]].

\bibitem{Kadota:2009sf}
  K.~Kadota and J.~Shao,
  Phys.\ Rev.\  D {\bf 80} (2009) 115004
  [arXiv:0910.5517 [hep-ph]].

\bibitem{Abada:2010kj}
  A.~Abada, A.~J.~R.~Figueiredo, J.~C.~Romao and A.~M.~Teixeira,
  JHEP {\bf 1010}, 104 (2010)
  [arXiv:1007.4833 [hep-ph]].

%
\bibitem{Allanach:2008ib}
  B.~C.~Allanach, J.~P.~Conlon and C.~G.~Lester,
  Phys.\ Rev.\  D {\bf 77}, 076006 (2008)
  [arXiv:0801.3666 [hep-ph]].




\bibitem{Casas:2001sr}
  J.~A.~Casas and A.~Ibarra,
  Nucl.\ Phys.\  B {\bf 618}, 171 (2001)
  [arXiv:hep-ph/0103065].

\bibitem{Martin:1993zk}
  S.~P.~Martin and M.~T.~Vaughn,
  Phys.\ Rev.\  D {\bf 50}, 2282 (1994)
  [Erratum-ibid.\  D {\bf 78}, 039903 (2008)]
  [arXiv:hep-ph/9311340].

\bibitem{Fonseca:2011vn}
  R.~Fonseca, M.~Malinsky, W.~Porod and F.~Staub,
  arXiv:1107.2670 [hep-ph].




\bibitem{Porod:2003um}
  W.~Porod,
  Comput.\ Phys.\ Commun.\  {\bf 153} (2003) 275
  [arXiv:hep-ph/0301101].

\bibitem{Porod:2011nf}
  W.~Porod and F.~Staub,
  arXiv:1104.1573 [hep-ph].

\bibitem{hep-ph/0202074} 
  P.~F.~Harrison, D.~H.~Perkins and W.~G.~Scott,
  Phys.\ Lett.\ B\ {\bf 530}, 167  (2002)
  [hep-ph/0202074].

\bibitem{arXiv:1111.0183} 
  M.~Khabibullin and f.~t.~T.~Collaboration,
  arXiv:1111.0183 [hep-ex].

\bibitem{DC2011}
Double CHOOZ experiment, talk by H.De. Kerrect at LowNu2011,
http://workshop.kias.re.kr/lownu11/

\bibitem{Bachacou:1999zb}
  H.~Bachacou, I.~Hinchliffe and F.~E.~Paige,
  Phys.\ Rev.\  D {\bf 62} (2000) 015009
  [arXiv:hep-ph/9907518].

\bibitem{Allanach:2000kt}
  B.~C.~Allanach, C.~G.~Lester, M.~A.~Parker and B.~R.~Webber,
  JHEP {\bf 0009} (2000) 004
  [arXiv:hep-ph/0007009].

\bibitem{Lester:2001zx}
  C.~G.~Lester,
  ``Model independent sparticle mass measurements at ATLAS''; 
  CERN-THESIS-2004-003

\bibitem{arXiv:1109.2352} 
  S.~Chatrchyan {\it et al.} [CMS Collaboration],
  Phys. Rev. Lett. {\bf 107} (2011) 221804; arXiv:1109.2352 [hep-ex].

\bibitem{arXiv:1109.6572} 
  G.~Aad {\it et al.} [ATLAS Collaboration],
  arXiv:1109.6572 [hep-ex].

\bibitem{Allanach:2002nj}
B.~C.~Allanach {\it et al.},
in {\it Proc. of the APS/DPF/DPB Summer Study on the Future of 
Particle Physics (Snowmass 2001) } ed. N.~Graf,
Eur.\ Phys.\ J.\  C {\bf 25}, 113 (2002)
[arXiv:hep-ph/0202233].

\bibitem{arXiv:1004.1092} 
  G.~Belanger, F.~Boudjema, P.~Brun, A.~Pukhov, S.~Rosier-Lees, 
  P.~Salati and A.~Semenov,
  Comput.\ Phys.\ Commun.\ \ {\bf 182}, 842  (2011)
  [arXiv:1004.1092 [hep-ph]].

\bibitem{arXiv:0803.2360} 
  G.~Belanger, F.~Boudjema, A.~Pukhov and A.~Semenov,
  Comput.\ Phys.\ Commun.\ \ {\bf 180}, 747  (2009)
  [arXiv:0803.2360 [hep-ph]].

\bibitem{hep-ph/0607059} 
  G.~Belanger, F.~Boudjema, A.~Pukhov and A.~Semenov,
  Comput.\ Phys.\ Commun.\ \ {\bf 176}, 367  (2007)
  [hep-ph/0607059].

\bibitem{hep-ph/0405253} 
  G.~Belanger, F.~Boudjema, A.~Pukhov and A.~Semenov,
  Comput.\ Phys.\ Commun.\ \ {\bf 174}, 577  (2006)
  [hep-ph/0405253].


\bibitem{arXiv:1107.5547} 
  J.~Adam {\it et al.} [MEG Collaboration],
  Phys.\ Rev.\ Lett.\ \ {\bf 107}, 171801  (2011)
  [arXiv:1107.5547 [hep-ex]].

\bibitem{arXiv:1103.0734} 
  T.~Schwetz, M.~Tortola and J.~W.~F.~Valle,
  New J.\ Phys.\ \ {\bf 13}, 063004  (2011)
  [arXiv:1103.0734 [hep-ph]].

\bibitem{hep-ph/0406166} 
  B.~C.~Allanach, A.~Djouadi, J.~L.~Kneur, W.~Porod and P.~Slavich,
  JHEP\ {\bf 0409}, 044  (2004)
  [hep-ph/0406166].


\end{thebibliography}
\end{document}